\documentclass[apjl]{emulateapj}
\usepackage{comment}
\usepackage{ifthen}
\usepackage{amsmath}
\usepackage{amsfonts}
\usepackage{amssymb}
\usepackage{multirow}
\usepackage{wasysym}
\usepackage{subfigure}
\usepackage{epsfig}
\usepackage[%
	colorlinks=true,	
	urlcolor=blue,		
	pdfpagelabels=true,
	hypertexnames=true,
	plainpages=false,
	naturalnames=true,
	pdftitle={Priors for Orbital Eccentricity},    
	pdfauthor={David M. Kipping},     
	linkcolor=red,          
	citecolor=green,        
        ]{hyperref}
\newcommand{\ms}{$\mathrm{m}\,\mathrm{s}^{-1}$}
\newcommand{\kms}{$\mathrm{km}\,\mathrm{s}^{-1}$}
\newcommand{\luna}{{\tt LUNA}}
\newcommand{\multi}{{\sc MultiNest}}
\newcommand{\cofiam}{{\tt CoFiAM}}
\newcommand{\blender}{{\tt BLENDER}}
\newcommand{\eccsamples}{{\tt ECCSAMPLES}}
\newcommand{\kepler}{\emph{Kepler}}

\newcommand{\kic}{KIC-8800954}
\newcommand{\koi}{KOI-1274}
\newcommand{\kep}{Kepler-421}
\newcommand{\koix}{KOI-1274.01}
\newcommand{\kepx}{Kepler-421b}

\newboolean{emulateapj}
\setboolean{emulateapj}{true}

\newboolean{astroph}
\setboolean{astroph}{true}

\newboolean{png}
\setboolean{png}{false}

\newboolean{color}
\setboolean{color}{true}

\shortauthors{Kipping et al.}
\shorttitle{Discovery of a Transiting Planet Near the Snow-Line}
\ifthenelse{\boolean{emulateapj}}{
    \newcommand{\titledag}{$\dagger$}
}{
    \newcommand{\titledag}{\dagger}
}

\begin{document}

\title {Discovery of a Transiting Planet Near the Snow-Line
\altaffilmark{\titledag}}

\author{
	{\bf	D.~M.~Kipping\altaffilmark{1,2},
		G.~Torres\altaffilmark{1},
		L.~A.~Buchhave\altaffilmark{1,3},
		S.~J.~Kenyon\altaffilmark{1},
		C.~E.~Henze\altaffilmark{4},
		H.~Isaacson\altaffilmark{5},
		R.~Kolbl\altaffilmark{5},
		G.~W.~Marcy\altaffilmark{5},
		S.~T.~Bryson\altaffilmark{4},
		K.~G.~Stassun\altaffilmark{6,7},
		F.~A.~Bastien\altaffilmark{6}
	}
}
\altaffiltext{1}{Harvard-Smithsonian Center for Astrophysics,
		Cambridge, MA 02138, USA; email: dkipping@cfa.harvard.edu}

\altaffiltext{2}{NASA Carl Sagan Fellow}

\altaffiltext{3}{Centre for Star and Planet Formation, Natural History Museum of 
                 Denmark, University of Copenhagen, DK-1350 Copenhagen, Denmark}

\altaffiltext{4}{NASA Ames Research Center, Moffett Field, CA 94035}

\altaffiltext{5}{University of California, Berkeley, CA 94720}

\altaffiltext{6}{Dept. of Physics \& Astronomy, Vanderbilt University,
                 1807 Station B, Nashville, TN 37235}

\altaffiltext{7}{Physics Dept., Fisk University, 1000 17th Ave. N,
                 Nashville, TN 37208}

\altaffiltext{$\dagger$}{
Based on archival data of the \emph{Kepler} telescope. 
}


\begin{abstract}

In most theories of planet formation, the snow-line represents a boundary
between the emergence of the interior rocky planets and the exterior ice giants.
The wide separation of the snow-line makes the discovery of transiting worlds
challenging, yet transits would allow for detailed subsequent characterization. 
We present the discovery of \kepx, a Uranus-sized exoplanet 
transiting a G9/K0 dwarf once every $704.2$\,days in a near-circular orbit. 
Using public \kepler\ photometry, we demonstrate that the two observed transits
can be uniquely attributed to the $704.2$\,day period. Detailed light curve 
analysis with \blender\ validates the planetary nature of \kepx\ to 
$>4$\,$\sigma$ confidence. \kepx\ receives the same insolation as a body at 
$\sim2$\,AU in the Solar System and for a Uranian albedo would have an effective 
temperature of $\sim180$\,K. Using a time-dependent model for the protoplanetary 
disk, we estimate that \kepx's present semi-major axis was beyond the snow-line 
after $\sim3$\,Myr, indicating that \kepx\ may have formed at its observed 
location.

\end{abstract}



\section{INTRODUCTION}
\label{sec:intro}


Extrasolar planets which fortitously transit their host star allow for a range
of measurements generally inaccessible via other methods \citep{winn:2010}.
For example, transits allow one to measure the planetary radius and inclination 
\citep{charbonneau:2000}, the true (rather than minimum) planetary mass when
combined with radial velocities \citep{charbonneau:2000}, the transmission
spectrum of the atmosphere \citep{seager:2000}, planetary rings 
\citep{barnes:2004} and even companion moons \citep{kipping:2009a,
kipping:2009b}. These widely recognized advantages have led to a recent 
explosion of photometric surveys to find such worlds (e.g. 
\textit{WASP}, \citealt{street:2003}; \textit{HATNet}, \citealt{bakos:2004};
\textit{KELT}, \citealt{pepper:2007,pepper:2012}; 
\textit{CoRoT}, \citealt{baglin:2006}; \kepler, \citealt{borucki:2009}; 
\textit{TESS}, \citealt{ricker:2010}; \textit{PLATO}, \citealt{rauer:2013}).

Unfortunately, one intrinsic bias of transit surveys is to short-period planets
via the scaling $\sim P_P^{-5/3}$ \citep{beatty:2008}, which is evident from the
paucity of planets with periods greater than one year in even the long-staring
mission \kepler\ \citep{burke:2014}. Long-period giant planets may be found more
easily with radial velocities (RVs), since the RV amplitude scales as 
$\sim M_P P_P^{-1/3}$. For planets with masses $M_P\geq0.3$\,$M_J$ and periods 
$P_P<2000$\,d, \citet{cumming:2008} find that the occurrence rates of RV planets 
generally increases with orbital period via $\sim P_P^{1/4}$. Further, numerous 
independent studies strongly suggest smaller planets are far more numerous than 
larger worlds \citep{jiang:2010,mayor:2011,fressin:2013,petigura:2013}. 
Therefore, whilst a lack of transiting long-period Jupiters could be due to 
their low occurrence, it is somewhat surprising that the more common 
Neptune-like planet has not been found to be transiting at long-periods
by \kepler.

In this work, we find the first member of this missing class of planets,
located in a regime which has not been previously probed by any of the planet 
detection techniques. \kepx\ opens the door to conducting transit experiments on 
an entirely new class of objects. This cold planet sits near the snow-line and 
the chemistry, composition and dynamical environment are likely to be greatly 
different from previous studies limited to planets with temperatues $\gtrsim 
500$\,K. We discuss how we identified this object first via the \emph{Kepler 
Mission} photometry in \S\ref{sec:photometry}. Follow-up observations are 
discussed in \S\ref{sec:followup}, which aid in planet validation with \blender, 
as discussed in \S\ref{sec:validation}, where we demonstrate the planetary 
nature of \kepx. Detailed light curve fits and credible intervals for the system 
parameters are provided in \S\ref{sec:fits}. Finally, we discuss the 
implications of our finding in \S\ref{sec:discussion}.

\section{KEPLER PHOTOMETRY}
\label{sec:photometry}


\subsection{Original Identification}
\label{sub:identification}

\koix\ was first announced as a candidate planet by \citet{batalha:2013} using
\kepler\ photometry of the 13.354 magnitude (\kepler\ bandpass) target
star using quarters 1-6 (Q1-6). A single-transit dip was detected at 
$\mathrm{BJD}_{\mathrm{UTC}}$ 2,455,325.76764 with a reported duration of 
$16.7283$\,hours and a depth of $2908$\,ppm. Despite observing just a single 
event, the high signal-to-noise ratio of $103.7$ made for a clear detection.
With a single transit, it is conventionally not possible to estimate the orbital
period of a planetary candidate and indeed we note that no orbital period is
reported in \citet{batalha:2013}. 

In the expanded catalog presented in \citet{burke:2014}, \koix\ is designated 
with an orbital period of $361.614906$\,days, which is presumably an approximate 
estimate based upon the transit duration and stellar properties. This same value 
is reported on the \href{http://exoplanetarchive.ipac.caltech.edu/}{NASA 
Exoplanet Archive} as well \citep{akeson:2013}, but for the Q1-6 catalog. By the 
time of the Q1-8 NASA Exoplanet Archive catalog, the period estimate includes an 
uncertainty of $362\pm82$\,days. However, with the Q1-12 candidate list, \koix\ 
appears to have been removed as a candidate, presumably since the expected 
second transit had not been found. This remains the case in the Q1-16 catalog 
too.

Despite being removed as a KOI, this candidate was identified as a potential
exomoon-hosting target by the ``Hunt for Exomoons with Kepler'' (HEK) project.
The HEK project utilizes an \textit{automatic target selection} (TSA) algorithm, 
which downloads the cumulative NASA Exoplanet Archive catalog (which includes all
four catalogs; Q1-6, Q1-9, Q1-12, Q1-16) and estimates both the dynamical
capacity of each planetary candidate for harboring an exomoon and the expected
detectability (see \citealt{hek:2012,hek:2013} for details on TSA). It was
during the HEK project's standard analysis of this target, that we identified
this KOI as potentially being much longer period than that reported on the
NASA Exoplanet Archive.

\subsection{Data Acquisition}
\label{sub:dataacquisition}

We downloaded the publicly available \kepler\ data for \koix\ from
the \href{http://archive.stsci.edu/}{Mikulski Archive for Space Telescopes}
(MAST). The downloaded data were released as part of Data Release 23 and were
processed using Science Operations Center (SOC) Pipeline version 9.1. Only
long-cadence (LC) data were available for this target.

Inspecting the Presearch Data Conditioning (PDC) time series allowed us to
exclude the proposed period of $\sim362$\,days, with no evidence for
transits at, or around, $\mathrm{BJD}_{\mathrm{UTC}}$ 2,455,687.76764 (Q9).
This result is consistent with the removal of \koix\ from the candidate
planet list on the NASA Exoplanet Archive as of the Q1-12 catalog.
Scanning the full time series though, a second transit signal is easily 
identified in Q13 at $\sim$ $\mathrm{BJD}_{\mathrm{UTC}}$ 2,456,030.
The depth and duration of this transit display a remarkable similarity
to the first event (this similarity is discussed in \S\ref{sub:period}). On
this basis, we considered a candidate period for \koix\ to be $704.2324$\,days,
nearly double the original estimate.

\subsection{Data Selection}
\label{sub:dataselection}

To fit light curve models to the \kepler\ data, it is necessary
to first remove instrumental and stellar photometric variability which can
distort the transit light curve shape. We break this process up into two stages:
i) pre-detrending cleaning ii) long-term detrending. In what follows, each 
quarter is detrended independently.

\subsection{Pre-Detrending Cleaning}
\label{sub:cleaning}

The first step is to visually inspect each quarter and remove any exponential 
ramps, flare-like behaviors and instrumental discontinuities in the data. We 
make no attempt to correct these artefacts and simply exclude them from the 
photometry manually. We then remove the two observed transits using the our
candidate ephemeris and clean the data of 3\,$\sigma$ outliers from a moving 
median smoothing curve with a 20-point window. During this stage and what
follows, we use the PDC data.

\subsection{Detrending with \cofiam}
\label{sub:detrending}

For the data used in the final transit light curve fits in \S\ref{sec:fits},
it is necessary to also remove the remaining long-term trends in the time
series. These trends can be due to instrumental effects, such as focus drift,
or stellar effects, such as rotational modulations. For this task, data are 
detrended using the Cosine Filtering with Autocorrelation Minimization (\cofiam) 
algorithm. \cofiam\ was specifically developed to protect the shape of a 
transit light curve and we direct the reader to our previous work 
\citep{hek:2013} for a detailed description.

To summarize the key features of \cofiam, it is essentially a Fourier-based
method which removes periodicities occurring at timescales greater than the
known transit duration. This process ensures that the transit profile is
not distorted in frequency space. \cofiam\ does not directly attempt to remove 
high frequency noise, since the Fourier transform of trapezoidal-like light
curve contains significant high frequency power \citep{waldmann:2012}. \cofiam\ 
is able to attempt dozens of different harmonics (we cap the maximum at 30) and 
evaluate the autocorrelation at a pre-selected timescale (we use 30\,minutes) 
and then select the harmonic order which minimizes this autocorrelation, as 
quantified using the Durbin-Watson statistic. This ``Autocorrelation 
Minimization'' component of \cofiam\ provides optimized data for subsequent 
analysis. We choose a 200\,hour window around either side of the two transits in 
order to provide ample out-of-transit baseline. Over this region, we find 
Durbin-Watson statistics consistent with negligible autocorrelation (2.02463 and 
1.9682).

\subsection{Orbital Period Identification}
\label{sub:period}

Inspection of the two transits reveals remarkable similarity (see 
Figure~\ref{fig:folded}). The second transit aligns nearly perfectly with the
first, indicating that this event is indeed due to the same transiting object.
This is further verified later in \S\ref{sub:indivfits}. 

\begin{figure*}
\begin{center}
\includegraphics[width=16.8 cm]{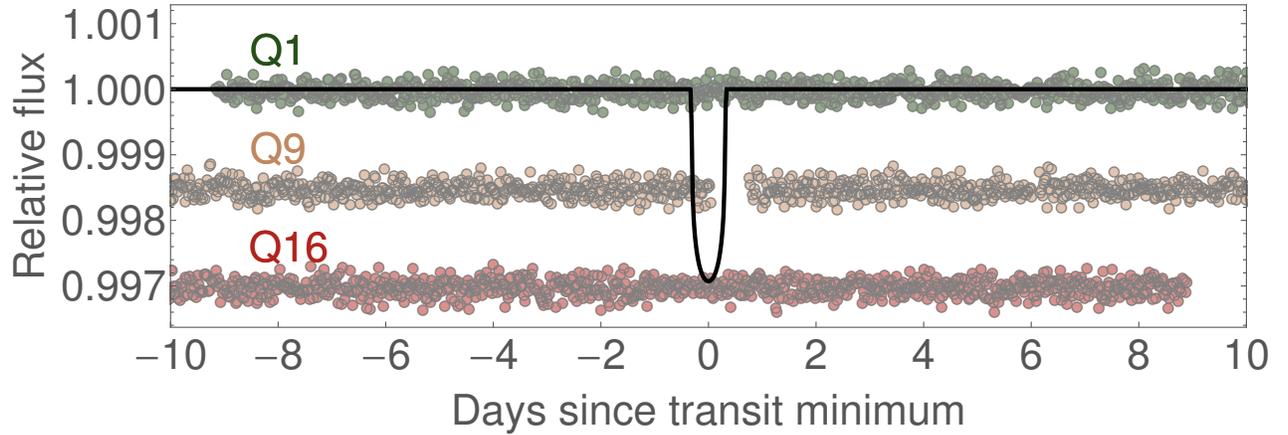}
\caption{
If the period of \koix\ were half that which we claim here, three additional 
transits would occur in the \kepler\ time series. For a linear ephemeris, the 
black line shows the expected location of such an event, which is highly 
inconsistent with the available data (circles).
}
\label{fig:missingtransit}
\end{center}
\end{figure*}

Despite these two events belonging to the same planet, the orbital period
remains ambiguous since other transits could be present in-between these two at 
integer fractions. The most easily misidentified period would be at half the 
putative period, implying $P=352.1$\,d, since this would require the fewest 
number of transits to have been missed. For a stictly linear ephemeris, this 
scenario requires three missing transits at $\mathrm{BJD}_{\mathrm{UTC}}$ 
2,454,973.668 in Q1, $\mathrm{BJD}_{\mathrm{UTC}}$ 2,455,677.866 in Q9 and 
$\mathrm{BJD}_{\mathrm{UTC}}$ 2,456,382.064 in Q16. As shown in 
Fig.~\ref{fig:missingtransit}, the \kepler\ archival data exhibit 
near-continuous temporal coverage around these times and clearly no additional
transits occur. It may be possible to conceive of very large amplitude and/or
finely tuned transit timing variations (TTVs) which could still hide the
missing transits, but we consider such a scenario highly contrived given the
available data.

In a similar vein, it is possible to exclude higher integer ratios. For example, 
if the true period were a third of our candidate value, four additional
transits would occur in the \kepler\ time series which again the excellent
temporal coverage excludes. Accordingly, we are able to demonstrate that our 
candidate orbital period for \koix, of $704.2$\,days, is correct.

\subsection{Stellar Rotational Period}
\label{sub:rotation}

Stars with active regions have a non-uniform surface brightness distribution,
which leads to time variable changes in brightness as the star rotates
\citep{budding:1977}. In practice, these active regions may evolve 
in location and amplitude over timescales of days to years, and even present 
evolving periodicities due to differential rotation \citep{reinhold:2013}. 
Despite this, the rotation period tends to produce a dominant peak in the 
frequency-domain allowing for an estimate of the rotation period using 
photometry alone \citep{basri:2011}.

Since the data are unevenly sampled and each quarter has a unique offset, we
elected to use a Lomb-Scargle style periodogram. The light curve model is a 
simple sinusoid and thus is linear with respect to the model parameters, for any 
trial rotation period, $P_{\mathrm{rot}}$. Using weighted linear least squares, 
we are guarenteed to find the global maximum likelihood solution at each
trial $P_{\mathrm{rot}}$. We scan in frequency space from twice the cadence
($\sim0.04$\,d) up to twice the total baseline of observations ($\sim2940$\,d)
making $10^5$ uniform steps in frequency. At each realization, we define the
``power'' as $(BIC_{\mathrm{null}}-BIC_{\mathrm{trial}})/BIC_{\mathrm{null}}$,
where BIC is the Bayesian Information Criterion and ``null'' and ``trial''
refer to the two models under comparison.

\begin{figure*}
\begin{center}
\includegraphics[width=18.0 cm]{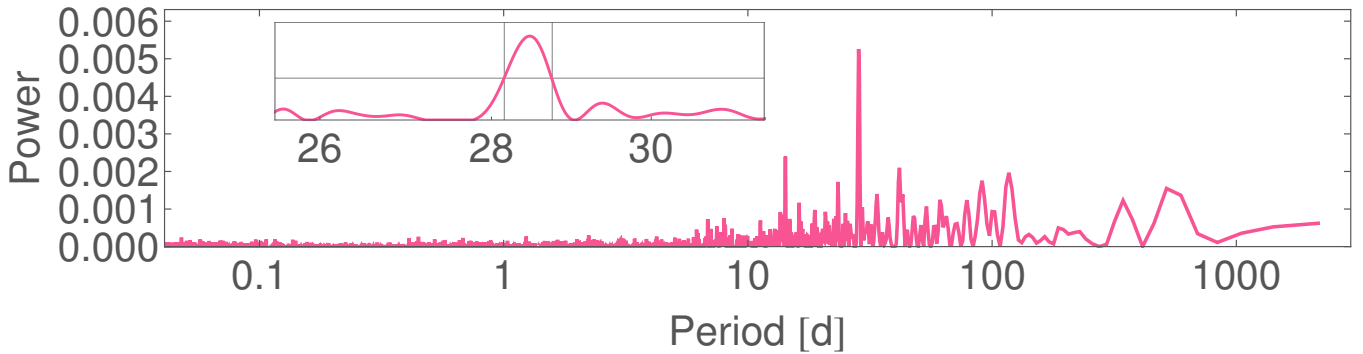}
\caption{Periodogram of the \kepler\ photometry for \kep. A zoomed-in
view of the peak at $P=28.5\pm0.3$\,days is shown in the top-left, which we
consider to be the stellar rotation period.}
\label{fig:periodogram}
\end{center}
\end{figure*}

The resulting periodogram, shown in Figure~\ref{fig:periodogram}, reveals a
strong peak at around 30\,days. Conducting a second periodogram with a finer
grid around this solution and defining the period uncertainty as the 
full width at half maximum, we find $P_{\mathrm{rot}} = 28.5\pm0.3$\,days. We 
note that this period lies close to the rotation period of typical spots on the
surface of the Sun of 27.3\,days. This information is used in 
\S\ref{sub:stellar} to constrain the age of \kep.

\begin{figure}
\begin{center}
\includegraphics[width=8.4 cm]{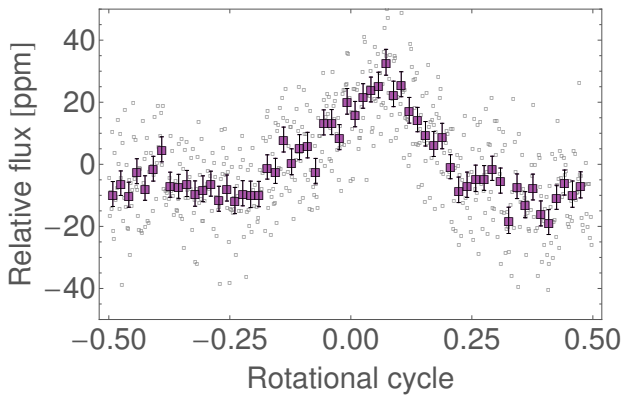}
\caption{Out-of-transit \kepler\ photometry for \kep\ folded upon our
maximum likelihood rotation period of $28.5$\,days. Gray points are
100-point phase binned, purple points are 1000-point phase binned.
}
\label{fig:stellar_folded}
\end{center}
\end{figure}

Folding the detrended \kepler\ photometry on our maximum likelihood rotation 
period reveals a quasi-sinusoidal pattern, with a peak-to-peak amplitude of 
$\sim40$\,ppm (see Figure~\ref{fig:stellar_folded}).

\section{FOLLOW-UP OBSERVATIONS}
\label{sec:followup}


\subsection{Overview}
\label{sub:followupoverview}

We describe here further observations and analyses we performed to both 
characterize the parent star, and to aid in addressing the probability that the 
photometric signal we detected might be due to an astrophysical false positive 
(i.e. other phenomena mimicking the transit) rather than a true planet around 
KOI-1274.

\subsection{Spectroscopy}
\label{sub:spectroscopy}

A high-resolution optical spectrum of \koi\ was obtained on 2011
August 4 using the fiber-fed echelle Spectrograph (FIES) on the 2.5\,m
Nordic Optical Telescope (NOT) on La Palma, Spain \citep{djupvik:2010}.
The resolving power delivered by the instrument with the medium
(1\farcs3) fiber setup is $R = 46,\!000$, and the 21-minute exposure
yielded an average signal-to-noise ratio per resolution element of 35
in the \ion{Mg}{1}\,b region (5190\,\AA).

On 2012 June 25 we acquired an additional high-resolution spectrum with the 
Keck~I Telescope on Mauna Kea (HI) and its HIRES spectrometer \citep{vogt:1994}. 
The exposure time was 10 min, and the size of the spectrograph slit was set to 
0\farcs86. Use of the standard setup and reduction procedures of the California 
Planet Search \citep{howard:2010,johnson:2010} resulted in a spectrum with 
$R \sim 60,\!000$ covering the range 3642--7990\,\AA, with a signal-to-noise
ratio per resolution element of 56 also in the \ion{Mg}{1}\,b region. The 
absolute radial velocity of the target was measured from this spectrum using 
telluric lines as the reference, and is $-25.4 \pm 0.1$\,\kms. We note that
this is compatible with that derived using the FIES spectrum, 
$-25.5 \pm 0.1$\,\kms.

We used the better-quality HIRES spectrum to search for signs of absorption 
lines from another star that might be causing the transit signal, by subtracting 
a spectrum closely matching that of the target star (after proper Doppler 
shifting and continuum normalization) and inspecting the residuals (Kolbl et al.
2014, in prep). The closest match (in a $\chi^2$ sense) was selected from a 
large library of observed spectra taken with the same instrument, spanning a 
wide range in temperature, surface gravity, and chemical composition. No 
evidence of secondary spectral lines was seen. To better quantify our 
sensitivity to faint companions, we subjected the residuals to a similar fitting 
process by injecting mock companions over a range of temperatures between 
3500\,K and 6000\,K, and with a wide range in relative velocities, and 
attempting to recover them. In this way we estimated we are sensitive to 
companions down to 1\% of the flux of the primary star, and velocity separations $\Delta RV$
greater than 10\,\kms. For smaller relative velocities the secondary lines would 
be blended with those of the primary and would not be detected.

\subsection{High-resolution imaging}
\label{sub:imaging}

\koi\ was observed as part of a large survey of \kepler\ Objects of
Interest by \cite{law:2013} using a robotic laser-guide-star adaptive
optics system known as Robo-AO, on the Palomar 60-inch telescope.
Images obtained on 2012 August 6 in the Sloan $i$ passband revealed a
close companion with a signal-to-noise ratio of 7, located at an
angular separation of $1\farcs10 \pm 0\farcs06$ in position angle
$241\arcdeg \pm 1\arcdeg$. This star is $3.75 \pm 0.44$ magnitudes
fainter than the target. \cite{law:2013} performed Monte Carlo
simulations in order to quantify the sensitivity to additional fainter
companions as a function of angular separation, by injecting and then
attempting to recover fake stars out to separations of 2\farcs5 for a
group of representative targets. For our \blender\ analysis below in
\S\ref{sec:validation} we have adopted their sensitivity curve
corresponding to observations of similar quality as \koi\ (see their
Figure~5, for medium performance).

\subsection{Centroid motion analysis}
\label{sub:centroids}

To search for false positives that may result, e.g., from a background eclipsing 
binary in the photometric aperture of \koi\ we measured the location of the 
transit signal relative to \koi\ via difference images formed by subtracting an 
average of in-transit pixel values from out-of-transit pixel values. If the 
transit signal is due to a stellar source, the difference image will show that 
stellar source, whose location is determined via Pixel Response Function (PRF) 
centroiding \citep{bryson:2013}. The centroid of an average out-of-transit image 
gives the location of \koi\ because the object is well isolated. The difference 
image centroid was compared to the out-of-transit image centroid, giving the 
location of the transit source relative to \koi. This was done for each of the 
two quarters in which a transit occurs. For further details of the procedure, 
see \cite{bryson:2013}.

In Quarter\,5 the measured transit location is offset from \koi\ by $0\farcs532 
\pm 0\farcs026$ with a position angle of about 316\arcdeg. In Quarter\,13 the 
transit location is offset by $0\farcs065 \pm 0\farcs027$ with a position angle 
of about 177\arcdeg. The uncertainties in these quarterly measurements are 
based on standard propagation of errors starting with the measured pixel-level 
noise. Clearly there is additional bias in each quarterly centroid measurement, 
so to determine the transit source location from these two measurements an 
average was computed as a least-squares constant fit to the quarterly offsets. 
The uncertainty of this average was computed both via standard propagation of 
errors for the fit and via a bootstrap estimate. We adopted the larger of these 
two uncertainty estimates (for details see \citet{bryson:2013}). The resulting 
average transit signal offset from \koi\ is $0\farcs248 \pm 0\farcs279$ with a 
position angle of about 308\arcdeg. This average position estimate is 
4.18\,$\sigma$ from the companion star at 1\farcs1 detected by \cite{law:2013}, 
ruling out that companion as the cause of the transit signal. Based on the 
1\,$\sigma$ uncertainty in this average we adopt a 3\,$\sigma$ radius of 
confusion of 0\farcs84, within which the centroid motion analysis is insensitive 
to the presence of contaminating stars (blends).

\subsection{Stellar properties}
\label{sub:stellar}

The spectroscopic parameters of \koi\ were derived from the
observations described in \S\ref{sub:spectroscopy} using the
Stellar Parameter Classification pipeline \citep[SPC;][]{buchhave:2012}.
Briefly, this algorithm cross-correlates the observed spectra against
a large library of calculated spectra based on model atmospheres by
R.\ L.\ Kurucz spanning a wide range of parameters, and assigns
stellar properties interpolated among those of the synthetic spectra
giving the best match. There is excellent agreement between the SPC
results from the individual spectra. The weighted mean of the two
estimates yielded $T_{\rm eff} = 5308 \pm 50$\,K, ${\rm [m/H]} = -0.25
\pm 0.08$~dex, $\log g = 4.61 \pm 0.10$~dex, and $v \sin I_{\star} = 0.0 \pm
0.5$\,\kms. These parameters correspond to a dwarf star of spectral
type G9 or K0.

We determined the mass and radius of the star, along with other
characteristics, by comparing these spectroscopic properties with
stellar evolution models from the Dartmouth series \citep{dotter:2008}
in a $\chi^2$ fashion. The procedure was analogous to that described
by \cite{torres:2008}. The results are listed in
Table~\ref{tab:stellarproperties}. The distance estimate of 320 pc
relies on an average $V$ magnitude of $13.56 \pm 0.04$
\citep{droege:2006,henden:2012,everett:2012} as well as our estimate of
the interstellar extinction toward \koi. We inferred the latter from
several sources \citep{hakkila:1997,schlegel:1998,drimmel:2003,amores:2005}
that yield an average reddening of $E(B-V) = 0.036 \pm 0.022$
(conservative error), corresponding to $A_V = 0.11 \pm 0.07$.

The uncertainty in the spectroscopic surface gravity is large enough
that the age of the star is essentially unconstrained by the
models. However, in \S\ref{sub:rotation} we describe the detection
of a rotation signature in the \kepler\ photometry with a period
of $P_{\rm rot} = 28.5 \pm 0.3$ days, which enables us to infer an age
using gyrochronology relations. We obtain consistent estimates of
$3.7_{-0.5}^{+0.9}$\,Gyr \citep{barnes:2007}, $4.1_{-0.6}^{+1.0}$\,Gyr
\citep{mamajek:2008}, and $4.3_{-0.7}^{+2.9}$\,Gyr \citep{meibom:2009}, in
which we have adopted a dereddened $B-V$ color of $0.73 \pm 0.04$ for
the star (from the indices derived by \citealt{henden:2012} and
\citealt{everett:2012}, and the above reddening value).

\begin{deluxetable}{lc}
\tablewidth{0pc}
\tablecaption{Stellar properties of \koi.\label{tab:stellarproperties}}
\tablehead{
\colhead{~~~~~~~~Property~~~~~~~~} & \colhead{Value}}
\startdata
$T_{\rm eff}$ (K)\dotfill         & $5308 \pm 50$ \\
$\log g$ (dex)\dotfill            & $4.61 \pm 0.10$ \\
${\rm [Fe/H]}$ (dex)\dotfill              & $-0.25 \pm 0.08$ \\
$v \sin I_{\star}$ (km~s$^{-1}$)\dotfill  & $0.0 \pm 0.5$ \\
$M_{\star}$ ($M_{\odot}$)\dotfill  & $0.794 \pm 0.030$ \\
$R_{\star}$ ($R_{\odot}$)\dotfill  & $0.757 \pm 0.029$ \\
$L_{\star}$ ($L_{\odot}$)\dotfill  & $0.40 \pm 0.06$ \\
$M_V$ (mag)\dotfill               & $5.90 \pm 0.12$ \\
Distance (pc)\dotfill             & $320 \pm 20$ \\
Age (Gyr)\dotfill                &  $4.0 \pm 0.8$
\enddata
\end{deluxetable}

\section{STATISTICAL VALIDATION}
\label{sec:validation}

\subsection{Overview}

A typical mass for a planet the size of \koix\
($\sim$4\,$R_{\oplus}$) is expected to be 10--20\,$M_{\oplus}$, based
on the range of properties of known exoplanets. The Doppler signal
induced on the host star would then have a semi-amplitude of only
0.8--1.6\,\ms, making it very challenging to detect with current
instrumentation around such a faint star ($V = 13.58$), particularly
given the long orbital period.  Dynamical ``confirmation'' in the
usual sense (by establishing that the orbiting object is of planetary
mass) is therefore not currently possible in this case. Instead, we
describe here our efforts to ``validate'' it statistically, by showing
that the likelihood of a true planet around \koi\ is orders of
magnitude larger than that of a false positive.

The main type of false positive that can mimic the transit signal in
this case involves blends with another eclipsing object in the
photometric aperture of \kepler. This includes background eclipsing
binaries (`BEBs'), background or foreground stars transited by a
(larger) planet (`BP'), or physically associated stellar companions
transited by another star or by a planet. The companions in these
latter cases are generally close enough to the target that they cannot
be resolved with high-resolution imaging. We refer to these
hierarchical triple configurations as `HTS' and `HTP', respectively,
depending on whether the orbiting object is a star or a planet. In
each of the four scenarios described above the eclipses are attenuated
by the light of the target and can be reduced to planetary
proportions, leading to confusion. Other types of false positives not
involving contamination by another object include grazing eclipsing
binaries, and transits of a small star in front of a giant star. Each
of these can easily ruled out because their signals would be
inconsistent with the significant second-to-third contact transit duration
($T_{23} = 14.18_{-0.15}^{+0.12}$\,hours; see \S\ref{sub:globalfits}) and
the measured properties of the star ($\log g = 4.61\pm0.10$\,dex).

To address the blends we relied on the \blender\ technique
\citep{torres:2004,torres:2011,fressin:2012}, which uses the detailed
shape of the transit light curve to weed out scenarios that lead to
the wrong shape for a transit. \blender\ simulates large numbers of
false positives over a wide range of stellar (or planetary)
parameters, and compares their synthetic light curves with the
\kepler\ photometry in a $\chi^2$ sense. Blends that provide poor fits
are considered to be excluded. This allows us to place tight
constraints on the types of objects composing the eclipsing pair that
yield viable blends, including their sizes or masses, as well as other
properties of the blends such as their overall brightness and color,
the linear distance between the background/foreground eclipsing pair
and the KOI, and even the eccentricities of the orbits. For details of
the procedure we refer the reader to the above sources, or recent
applications of \blender\ by \cite{borucki:2013}, \cite{meibom:2013}, and
\cite{ballard:2013}. Following the nomenclature in those studies, the
objects in the eclipsing pair are designated as the ``secondary'' and
``tertiary'', and the target itself is referred to as the ``primary''.
Secondary and tertiary stellar properties (masses, sizes, and absolute
brightness in the \kepler\ and other passbands) were drawn from model
isochrones from the Dartmouth series \citep{dotter:2008}, and the
properties adopted for the primary are those reported in
\S\ref{sub:stellar}, supplemented with others inferred from the
adopted isochrone. The long-cadence photometry we used for \koi\ was
processed and detrended as described earlier.

\begin{figure}
\epsscale{1.15} 
\plotone{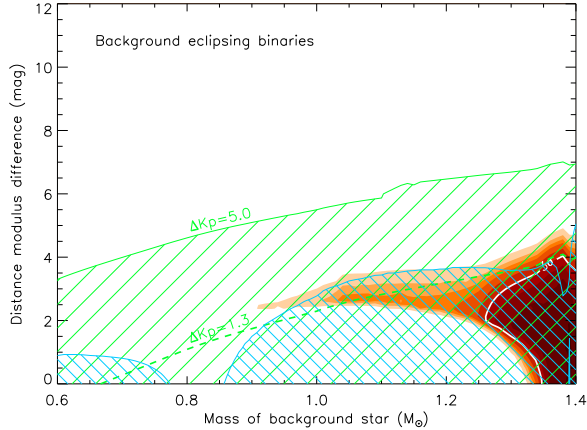}

\figcaption[]{Map of the $\chi^2$ surface (goodness of fit) for
  \koi.01 corresponding to blends involving background eclipsing
  binaries. On the vertitcal axis is represented the linear distance
  between the BEB and the target ($D_{\rm BEB} - D_{\rm targ}$),
  expressed for convenience in terms of the difference in distance
  modulus, $\Delta\delta = 5\log(D_{\rm BEB}/D_{\rm targ}$). Only
  blends within the solid white contour (darker colors) provide fits
  to the \kepler\ light curve that are within acceptable limits
  \citep[3$\sigma$, where $\sigma$ is the significance level of the
    $\chi^2$ difference compared to a transiting planet model fit;
    see][]{fressin:2012}. Other concentric colored areas (lighter
  colors) represent fits that are increasingly worse (4$\sigma$,
  5$\sigma$, etc.), which we consider to be ruled out. The blue
  cross-hatched areas correspond to regions of parameter space where
  the blends are either too red (left) or too blue (right) compared to
  the measured $r-K_s$ color of the target, by more than three times
  the measurement uncertainty. The dashed green line labeled $\Delta
  K\!p = 1.3$ is tangent to the white contour from above and
  corresponds to the faintest viable blends. The green line labeled
  $\Delta K\!p = 5.0$ represents the spectroscopic limit on faint
  background stars. All simulated blends below this line (hatched
  region) are brighter and are generally excluded if the BEB is
  angularly close enough to the target to fall within the slit of the
  spectrograph (see text). Thus, very few blends remain viable.
\label{fig:bs}}

\end{figure}

\subsection{Blend Frequency}
\label{sub:blendprior}

Our simulations with \blender\ rule out all scenarios in which a pair
of eclipsing stars orbits the target (HTS). The resulting light curves
invariably have the wrong shape for the transit, or produce secondary
eclipses that are not observed in the photometry of \koi. Background
eclipsing binaries with two stellar components are only able to
produce viable false positives if the secondaries are restricted to a
very narrow range of masses between about 1.25\,$M_{\odot}$ and
1.43\,$M_{\odot}$, as well as a limited interval in brightness ($K\!p$
magnitude) relative to the primary ($\Delta K\!p \lesssim 1.3$). We
illustrate this in Figure~\ref{fig:bs}, which shows the $\chi^2$
landscape for all blends of this kind in a representative
cross-section of parameter space. Regions outside of the 3-$\sigma$
contour correspond to configurations with light curves that give poor
fits to the \kepler\ photometry, i.e., much worse than a true planet
fit. These scenarios are therefore excluded. Other constraints
available for \koi\ place further restrictions on viable blends. In
particular, comparing the $r-K_s$ colors of the simulated blends with
the measured color index of the target \citep[$r-K_s = 1.779 \pm
  0.023$;][]{brown:2011}, we find that most of the BEB scenarios
allowed by \blender\ are too blue by more than 3$\sigma$, and are
therefore also excluded.  Additionally, the analysis of our Keck/HIRES
spectrum generally rules out companions within 5 magnitudes of the
primary if they are closer than 0\farcs43 (half-width of the
spectrograph slit) and their radial velocity (RV) is offset by more
than 10\,\kms\ from that of the primary. If $\Delta RV <
10$\,\kms\ line blending could prevent their detection, so those stars
are not necessarily excluded. Stars brighter than $\Delta K\!p = 5$
but outside of 0\farcs43 are only ruled out if they are above the
detection threshold from the high-resolution imaging, which is a
function of their angular separation. We show these two observational
constraints in Figure~\ref{fig:bs}. Other constraints are discussed
below.

\begin{figure}
\epsscale{1.15}
\plotone{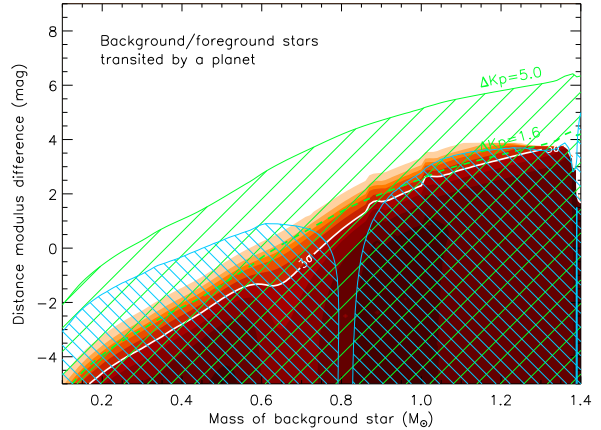}

\figcaption[]{Similar to Figure~\ref{fig:bs} (and with the same color
  scheme) for blends involving background or foreground stars
  transited by a planet (BP). The faintest blends giving acceptable
  fits have $\Delta K\!p = 1.6$ relative to the target (dashed green
  line), and are all excluded by the spectroscopic constraint ($\Delta
  K\!p = 5.0$, hatched green area) unless the intruding star is more
  than 0\farcs43 from the target, or within 0\farcs43 but with $\Delta
  RV < 10$\,\kms\ (see text).\label{fig:bp}}

\end{figure}

For blend scenarios involving a background or foreground star
transited by a planet (BP) there is a wide range of secondary masses
that yield acceptable fits to the \kepler\ light curve, as shown in
Figure~\ref{fig:bp}.  The faintest of these blends are 1.6 mag fainter
than the primary in the $K\!p$ band.  However, as in the case of BEBs,
the observational constraints severely limit the pool of viable false
positives. In particular, many of them are either too red or too blue
compared to the measured color index of the target, or are bright
enough that they would generally have been detected spectroscopically
(but not always; see above).

\begin{figure}
\epsscale{1.15}
\plotone{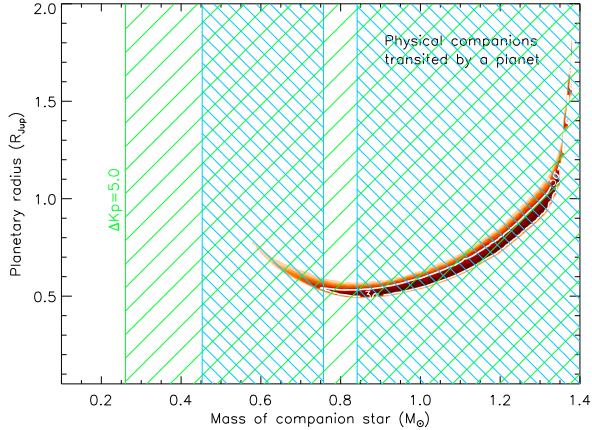}

\figcaption[]{Similar to Figure~\ref{fig:bs} for the case of physical
  companions to \koi\ that are transited by a planet (HTP). Only
  companion stars with masses between about 0.75\,$M_{\odot}$ and
  0.84\,$M_{\odot}$ (i.e., very similar to the target star itself)
  yield blend colors that are consistent with the measured $r-K_s$
  index of \koi. However, these blends are generally eliminated by the
  spectroscopic consraint (but not always; see text), as they are
  brighter than $\Delta K\!p = 5.0$.\label{fig:htp}}

\end{figure}

The $\chi^2$ map for scenarios involving a physically associated
companion to \koi\ that is transited by a larger planet (HTP) appears
in Figure~\ref{fig:htp}, and shows the size of the planetary tertiary
that can mimic the transit signal as a function of the companion
star. \blender\ restricts viable blends to be in a very narrow strip
of parameter space corresponding to secondary masses between about
0.75\,$M_{\odot}$ and 1.35\,$M_{\odot}$, and planetary sizes
0.5--1.2\,$R_{\rm Jup}$ (5.5--13.5\,$R_{\oplus}$). As before, color
and brightness constraints permit us to exclude most of these blends
(see Figure~\ref{fig:htp}), but not all (e.g., not ones within
0\farcs43 where the spectral lines of the companion and the primary
would be blended).

The frequencies with which each of the blend configurations are
expected to occur were computed via Monte Carlo experiments, in which
we simulated large numbers of blends and rejected those that give poor
fits to the transit photometry or that would have been detected with
the aid of our follow-up observations. We then counted the remaining
blends to derive their frequencies. These experiments relied on a
number of ingredients including the known distributions of binary star
properties, the number density of stars near the target, and the rates
of occurrence of transiting planets and eclipsing binaries inferred
from the \kepler\ observations themselves. Those rates (and any
dependence they may have on orbital period or other properties) are
conveniently implicit in the KOI lists generated by the \kepler\ team
and in the eclipsing binary catalog of \cite{slawson:2011}, when
normalized by the total number of targets observed by \kepler. We have
taken advantage of that information below.

For the HTP case we simulated companion stars following the
distributions of binary properties given by \cite{raghavan:2010} (mass
ratios, orbital periods, eccentricities), and placed them in random
orbits around the primary and at random orbital phases. We derived
their relevant stellar properties (size, brightness, colors) from the
isochrone used for the target. We assigned to each of these companions
a random transiting planet drawn from the KOI list hosted at the NASA
Exoplanet Archive\footnote{\tt
  http://exoplanetarchive.ipac.caltech.edu/} (downloaded 2014 March
26), but accepted as viable only those with periods similar to that of
\koi.01 (within a factor of two). The rationale for this is that the
relevant blend frequency is that of configurations involving planets
with periods near that of the candidate, since those frequencies
depend strongly on period.  We then examined the properties of the
companion stars and their planets, and rejected configurations that do
not satisfy the \blender\ restrictions on companion mass, planetary
size, and orbital eccentricity as illustrated in
Figure~\ref{fig:htp}. We further rejected those that would have been
detected in our high-resolution imaging, spectroscopy, centroid motion
analysis, or by their colors. In applying the spectroscopic constraint
we discarded blends brighter than $\Delta K\!p = 5.0$ mag if the
companion star is within 0\farcs43 of the target, unless its radial
velocity computed from the simulated orbit around the target is within
10\,\kms\ of that of \koi. In that case the spectral lines would
be heavily blended with those of the primary and might be missed, so
we consider those blends still viable.  Finally, we retained only
configurations that are dynamically stable according to the criterion
of \cite{holman:1999}.  We repeated this experiment a large number of
times, counting the surviving blends and finally multiplying by the
46\% frequency of non-single stars from \cite{raghavan:2010}. The
resulting HTP blend frequency is very small, $4.9 \times 10^{-11}$,
which is due to a combination of very strong observational
constraints, the well-defined shape of the transit, and the rare
occurrence of larger transiting planets in such wide orbits that can
be involved in blends.

The calculation of the frequency of background or foreground stars
transited by a planet (BP) proceeded in a similar fashion, and depends
on the number density of stars in the vicinity of \koi\ (stars per
square degree). Using the Galactic structure model of \cite{robin:2003},
we began by generating a list of simulated stars in a 5 square-degree
area around the target, including their kinematic properties (radial
velocity). We then drew stars randomly from this list assigning them a
random angular separation from the target within the 0\farcs84,
3-$\sigma$ exclusion radius from our centroid motion analysis (since
stars outside of this area would have been detected).  To each of
these stars we assigned a transiting planet from the KOI list,
retaining only those within a factor of two of the period of \koi.01,
as before. Blends involving background/foreground stars and their
planets that do not meet the \blender\ constraints were rejected,
along with those that would have been flagged by our follow-up observations 
(imaging, spectroscopy, and color). We retained configurations in which the 
velocity of the simulated secondaries are within 10\,\kms\ of that measured for 
\koi\ ($-24.7$\,\kms; Sect. 3.1), which would be spectroscopically undetectable. 
The resulting frequency of false positives of this kind, after normalizing by 
the ratio of areas between the centroid exclusion region and 5 square degrees, 
is only $3.5 \times 10^{-14}$. The contribution of these kinds of blends to the 
overall frequency is therefore negligible, and as before this is partly a 
consequence of how uncommon long-period transiting planets are, which in turn 
has to do with the low probability of transit.

To assess the frequency of BEBs acting as blends we used the Galactic structure 
model of \cite{robin:2003} as above, drawing secondaries and assigning a 
companion star (tertiary) from the distributions of binary properties of 
\cite{raghavan:2010}. We assigned periods to these blends drawn randomly from 
the list of \kepler\ binaries by \cite{slawson:2011}, and kept only those within 
a factor of two of the period of \koi.01. After filtering out BEBs with 
properties that make them inviable according to \blender, we applied the 
observational constraints in the same way as for the BP scenario, and counted 
the surviving blends. Their frequency in this case is so small that we can only 
place an upper limit of $\sim10^{-14}$. In this case the smallness of the number 
is related to the low frequency of eclipsing binaries with periods as long as 
704\,days (again largely a result of the low probability of eclipse).

The total blend frequency is the sum of the HTP, BP, and BEB
contributions, or $4.9 \times 10^{-11}$. While this is a very small
figure, the \emph{a priori} frequency of a true transiting planet such
as \koix (`planet prior') is also expected to be very small. Our aim
in the present work is to validate the candidate to a very high level
of confidence equivalent to a 3$\sigma$ significance, consistent with
previous applications of \blender. For this we require a planet prior
that is at least $1/(1-99.73\%) \approx 370$ times larger than the
total blend frequency, or $\sim 2 \times 10^{-8}$. 

\subsection{Planet Prior}
\label{sub:planetprior}

We here discuss our estimation of the \emph{a priori} probability of \kepler\ 
detecting a genuine planet similar to \koix- the planet prior. We define the 
planet prior as the prior probability of a star having a planet similar to 
\koix\ (the occurrence rate) multiplied by the prior probability of \kepler\ 
detecting such a planet in transit (the detection probability). We discuss each 
of these in what follows.

\subsubsection{Occurrence rate calculation}

The occurrence rate
of a planet at a specific orbital period and size cannot be reasonably defined,
since it will always be infinitessimal. Instead, one must consider a confidence 
interval defining ``similar'' planets to the one in question, and then integrate 
the occurrence rate distribution over this range. For consistency with previous 
\blender\ works, we define the period interval as $P/2\to2P$, where
$P$ is the orbital period equal to 704.1984\,d (see
\S\ref{sec:fits}). Similarly, previous \blender\ works have defined the 
size interval to be the 3\,$\sigma$ confidence region of $R_P$, which for \koix\
is $3.55$\,$R_{\oplus}<R_P<5.03$\,$R_{\oplus}$ (see \S\ref{sec:fits}).

Unlike previous \blender\ analyses \citep{torres:2011, fressin:2011,
fressin:2012}, we are unable to use the \kepler\ statistics themselves 
\citep{fressin:2013} to define the occurrence rate, since these are only 
complete up to 418\,d. Instead, we turn to the radial velocity occurrence rates
from \citet{cumming:2008}, which are complete for $P<2000$\,d and 
$M_P\geq0.3$\,$M_J$. One complication introduced by this decision is that
we must convert radii to masses. Although empirical mass-radius relations have
been derived for $\sim4$\,$R_{\oplus}$ planets \citep{weiss:2014}, these 
relations are calibrated to planets which are closer to their host star than 
\koix. Instead, given that \koix\ most closely resembles Uranus in size at just 
2.5\% larger and both worlds are ``cold'', being $\lesssim200$\,K, we simply 
adopt the Uranian mean density (1.27\,g\,cm$^{-3}$) to estimate masses. This 
changes our 3\,$\sigma$ confidence interval from 
$3.55$\,$R_{\oplus}<R_P<5.03$\,$R_{\oplus}$ to 
$10.0$\,$M_{\oplus}<M_P<29.4$\,$M_{\oplus}$ (with the most likely value being
$16.0$\,$M_{\oplus}$).

This mass estimation process reveals that \koix\ likely has a mass below
0.3\,$M_J$ ($=95$\,$M_{\oplus}$), which is a problem given that the 
\citet{cumming:2008} relation only extends down to 0.3\,$M_J$. However,
we also know that empirically determined occurrence rates consistently show
that smaller planets outnumber their bigger brothers \citep{jiang:2010}. We
may therefore simply push our mass range up to the 0.3\,$M_J$ boundary and
any derived occurence rate will be a conservative estimate. This ``push'' could
be done by multiplying the mass range until the lower limit equals 0.3\,$M_J$
giving $0.30$\,$M_J<M_P<0.85$\,$M_J$. Alternatively, one may simply add a
constant to both limits until the lower limit equals 0.3\,$M_J$, giving
$0.30$\,$M_J<M_P<0.36$\,$M_J$. The narrow range of this latter estimate yields
a more conservative occurrence rate and so we adopt it from here on.

The \citet{cumming:2008} occurrence rate is described by a power-law function:

\begin{align}
\mathrm{d}N &= C M^{\alpha} P^{\beta}\,\mathrm{d}\log M\,\mathrm{d}\log P
\end{align}

\begin{figure}
\begin{center}
\includegraphics[width=8.4 cm]{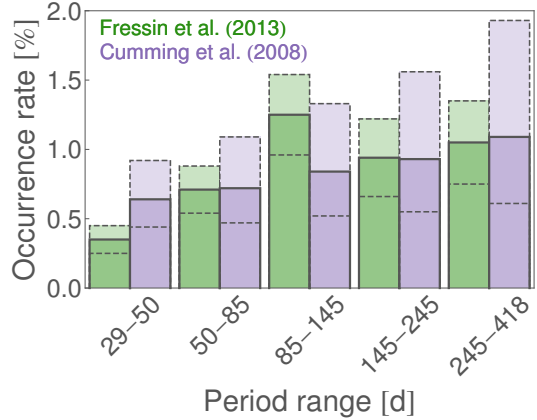}
\caption{
Occurrence rates of ``giant'' planets in five period bins from 
\citet{fressin:2013} (green bars) and \citet{cumming:2008} (purple bars).
Dashed lines mark the $\pm 1$\,$\sigma$ confidence interval.
The \citet{cumming:2008} values provide excellent agreement with those of
\citet{fressin:2013}, yet extend to longer periods up to 2000\,d, thereby
offering a viable method of computing the planet prior for \koix.
} 
\label{fig:occurrence}
\end{center}
\end{figure}

It is this function we must integrate between our defined interval to compute 
the occurrence rate. Before proceeding, we calculate some example occurrences 
using this law and the \citet{cumming:2008} estimates and associated 
uncertainties for $\alpha$ and $\beta$ (we assume they are normally 
distributed). We consider the range $0.3\to10$\,$M_J$ to define the 
range of ``giant planets''. This may be compared to the ``giants'' bin 
given by \citet{fressin:2013} of $6\to22$\,$R_{\oplus}$. Using the five longest
period intervals presented by \citet{fressin:2013}, we show the comparison of
these two occurrence rate estimates in Figure~\ref{fig:occurrence}.
The evident close agreement of these two estimates demonstrates the reliability 
of the \citet{cumming:2008} power-law. Accordingly, we use the law to finally 
estimate a conservative occurrence rate of $0.041_{-0.020}^{+0.038}$\% for 
planets similar to \koix.

\subsubsection{Detection probability calculation}

\koix\ presents a high signal-to-noise ratio (SNR) transit with 
$(R_P/R_{\star})^2=(2.508_{-0.058}^{+0.082})\times10^{-3}$. With a SNR exceeding 
40, the recovery rate of such signals is expected to be $\sim100$\% 
\citep{fressin:2013}. In light of this, the only relevant issue for the 
detection probability is the geometric transit probability of such a long-period 
planet. The geometric transit probability of a circular orbit planet is simply
$1/(a/R_{\star})$ but circularity cannot be reasonably assumed for a long-period 
planet such as \koix. If one assumes the eccentricity distribution may be
described by a Beta distribution \citep{beta:2013}, then the
eccentricity-marginalized transit probability is derived by 
\citet{ECCSAMPLES:2014} to be

\begin{align}
\mathrm{P}(\mathrm{transit}|a/R_{\star}) &= \Big(\frac{1}{a/R_{\star}}\Big) \Big(\frac{\Gamma[\alpha_e+\beta_e]}{\beta_e-1} \Big) \gamma_1,\\
\gamma_1 &= \,_2\tilde{F}_1[1,\alpha_e;\alpha_e+\beta_e-1;-1],
\end{align}

where $\alpha_e$ and $\beta_e$ are the Beta distribution shape parameters
and $_2\tilde{F}_1$ is Gauss' hypergeometric function. To compute this 
probability, we draw samples from the joint posterior distribution of
$\alpha_e$ and $\beta_e$ for the long-period sample described in 
\citet{beta:2013} and samples for $(a/R_{\star})$ from our light curve fits
(see \S\ref{sec:fits}). This yields $\mathrm{P}(\mathrm{transit}|a/R_{\star}) = 
0.341_{-0.017}^{+0.033}$\%.

\subsection{Final Result}
\label{sub:validationfinal}

Using the results of \S\ref{sub:planetprior}, we may multiply the planet 
occurrence rate prior by the transit probability prior to estimate our final 
planet prior of $(1.41_{-0.69}^{+1.32})\times10^{-6}$. Recall in 
\S\ref{sub:blendprior} that any planet prior above $\sim2\times10^{-8}$ would
indicate a $>3$\,$\sigma$ statistical validation of \koix. In fact, in this case 
the probability of a planetary nature for KOI-1274.01 is approximately 28,000 
greater than that of a false positive, corresponding to a confidence level for 
the validation of 4.1$\sigma$. We therefore refer to \koix\ as \kepx\ throughout 
the remainder of this paper, and similarly refer to the host star as \kep.

\section{LIGHT CURVE FITS}
\label{sec:fits}


\subsection{Global Fit}
\label{sub:globalfits}

Here we here discuss our final transit light curve fit of both epochs (i.e. a 
global fit) from which we derive our final system parameters. In what follows, 
we use the detrended PDC \kepler\ photometric time series as our input data, for 
which details on the detrending are described in \S\ref{sub:detrending}. 

We model the transit light curve using the standard \citet{mandel:2002}
algorithm employing the quadratic limb darkening law. This simple transit model
assumes a spherical, opaque planet transiting a spherically symmetric luminous
star on a Keplerian orbit. We resample the long-cadence data into short-cadence
sampling following the method described in \citet{binning:2010}, to avoid 
smearing effects. Our model has 10 free parameters in total. These are the 
orbital period, $P$, the time of transit minimum, $\tau$, the ratio-of-radii, 
$p$, the mean stellar density, $\rho_{\star}$, the impact parameter, $b$, the 
orbital eccentricity, $e$, the argument of periapsis, $\omega$, the blend 
factor, $B$, and the quadratic limb darkening coefficients $q_1$ and $q_2$. All 
of these parameters have uniform priors in our fits, except $\rho_{\star}$ which 
uses a log-normal prior, $e$ which uses a Beta prior, $B$, which uses a normal 
prior and $\omega$ which uses a periodic uniform prior. Note that we do not fit 
directly for the standard quadratic limb darkening coefficients $u_1$ and $u_2$, 
but rather use the transformed parameters $q_1$ and $q_2$ as advocated in 
\citet{LDfitting:2013}, in order to impose efficient, uninformative and physical 
priors for the limb darkening profile.

In the case of $\rho_{\star}$, we use an informative log-normal prior rather
than an uninformative choice, as used for the other parameters. Using an 
informative prior in $\rho_{\star}$ allows for the orbital eccentricity to be 
constrained via Asterodensity Profiling (AP) \citep{AP:2014}, specifically via 
the photoeccentric effect \citep{dawson:2012}. Using a $\rho_{\star}$ prior from
asteroseismology, \citet{kepler22:2013} recently demonstrated this principle on 
the planet Kepler-22b. In the case of \kep, this dwarf star is too faint 
($K\!p=13.354$) for the detection of asteroseismic modes ($K\!p\lesssim12$) and 
thus we must find an alternative independent constraint on the stellar density.
Recently, \citet{flicker:2014} have proposed that brightness variability on an
8-hour timescale, so-called ``flicker'' \citep{bastien:2013}, may be used as an 
alternative constraint for fainter stars. Using the method described in
\citet{flicker:2014}, we estimate a flicker of $F_8 = (18.3 \pm 5.2)$\,ppm which
yields a constraint on the stellar density of 
$\log_{10}(\rho_{\star}\,\mathrm{[kg}\,\mathrm{m}^{-3}\mathrm{]}) = 
(3.08\pm0.27)$\,dex. As discussed in \citet{flicker:2014}, a flicker-based
estimate of the stellar density has a probability distribution well-described
by a log-normal distribution, which is why we use this function here.

In the case of $e$, we use a loose but informative Beta distribution prior
described by shape parameters $\alpha_e$ and $\beta_e$. The Beta distribution 
provides the closest match to observed distribution of eccentricities from
radial velocity surveys \citep{beta:2013}. We use the ``long-period'' sample 
$P\gtrsim1$\,year described in \citet{beta:2013}, described by $\alpha_e=1.12$
and $\beta_e=3.09$. Beta samples were computed on the fly using the \eccsamples\ 
algorithm \citep{ECCSAMPLES:2014}. Finally, for the blend prior, we make use of 
the Robo-AO constrast measurement of the nearby companion \citep{law:2013}, 
which allows us to account for the extra contaminating light which dilutes the 
transit depth. In addition to this, we use an extra fixed blend factor unique to 
each quarter due to contaminating light identified in the MAST database.

We also mention that the out-of-transit baseline flux for each transit epoch
is also fitted. However, in this case, we use a linear minimization for the
baseline flux, similar to that described by \citet{kundurthy:2011}. This treats
the baseline flux simply as a nuisance parameter which is not marginalized
against, but rather minimized at each Monte Carlo realization. This allows us
to reduce the number of free parameters yet retain just 10 model parameters.

\begin{figure*}
\begin{center}
\includegraphics[width=18.0 cm]{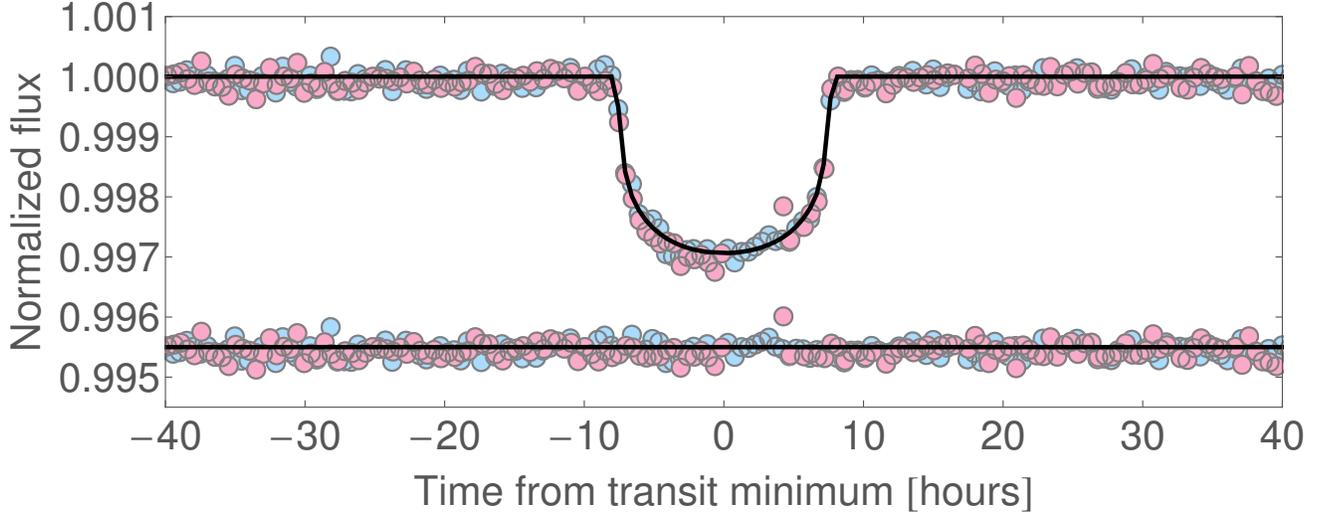}
\caption{Folded transit light curve. Blue points are associated with epoch 
1, and red points with epoch 2. Maximum a posteriori model fit is shown in
solid gray, with the corresponding (offset) residuals shown below.} 
\label{fig:folded}
\end{center}
\end{figure*}

To regress our 10-parameter model to the observations, we employed the 
mulitmodal nested sampling algorithm \multi\ described in 
\citet{feroz:2008,feroz:2009}. We use 4000 live points with constant efficiency
mode turned off and set an enlargement factor of 0.1. The maximum 
\emph{a posteriori} model parameters and their associated 68.3\% credible 
intervals are provided in Table~\ref{tab:finalparams}. We also show the folded 
transit light curve and the maximum \emph{a posteriori} transit model in 
Figure~\ref{fig:folded}.

\begin{table}
\caption{Final parameter estimates for \kepx.
$^{\dagger}$ = assuming a Bond albedo similar to Neptune/Uranus of 0.30.
$^{*}$ = equivalent semi-major axis of the \kepx\ if planet orbited the Sun
and received an insolation level of $S_{\mathrm{eff}}$.
} 
\centering 
\begin{tabular}{c c c c c c c} 
\hline\hline
Parameter & Estimate \\ [0.5ex] 
\hline
\textit{Fitted parameters} & & \\
\hline
$P$ \dotfill & $704.1984_{-0.0016}^{+0.0016}$ \\
$\tau$ [BKJD$_{\mathrm{UTC}}$-2,455,000] & $325.7674_{-0.0012}^{+0.0012}$ \\
$(R_P/R_{\star})$ \dotfill & $0.05008_{-0.00059}^{+0.00081}$ \\
$\rho_{\star}$\,[g\,cm$^{-3}$] \dotfill & $1.58_{-0.37}^{+0.22}$ \\
$b$ \dotfill & $0.21_{-0.14}^{+0.17}$ \\
$q_1$ \dotfill & $0.53_{-0.12}^{+0.16}$ \\
$q_2$ \dotfill & $0.311_{-0.095}^{+0.116}$ \\
$e$ \dotfill & $0.041_{-0.034}^{+0.095}$ \\
$\omega$\,[$^{\circ}$] \dotfill & $160_{-120}^{+150}$ \\
\hline
\textit{Other transit parameters} & & \\
\hline
$(a/R_{\star})$ \dotfill & $346_{-29}^{+16}$ \\
$i$\,[$^{\circ}$] \dotfill & $89.965_{-0.031}^{+0.024}$ \\
$u_1$ \dotfill & $0.814_{-0.069}^{+0.057}$ \\
$u_2$ \dotfill & $-0.08_{-0.12}^{+0.14}$ \\
$T_{14}$\,[hours] \dotfill & $15.79_{-0.10}^{+0.12}$ \\
$T_{23}$\,[hours] \dotfill & $14.18_{-0.15}^{+0.12}$ \\
$T_{12} \simeq T_{34}$\,[hours] \dotfill & $0.785_{-0.035}^{+0.105}$ \\
\hline
\textit{Physical parameters} & & \\
\hline
$R_P$\,[$R_{\oplus}$] \dotfill & $4.16_{-0.16}^{+0.19}$ \\
$a$\,[AU] \dotfill & $1.219_{-0.106}^{+0.089}$ \\
$T_{\mathrm{eq}}^{\dagger}$\,[K] \dotfill & $184.8_{-4.8}^{+8.6}$ \\
$S_{\mathrm{eff}}$\,[$S_{\oplus}$] \dotfill & $0.276_{-0.028}^{+0.055}$ \\
$a_{\mathrm{eff}}^{*}$\,[AU] \dotfill & $1.90_{-0.17}^{+0.10}$ \\ [1ex]
\hline 
\end{tabular}
\label{tab:finalparams} 
\end{table}

\subsection{Individual Fits}
\label{sub:indivfits}

We also attempted independent fits of each of our two individual transits. The
purpose of the these fits was to verify quantitatively that the two transits are 
consistent with being due to the same transiting body (see \S\ref{sub:period}). 
In these fits, it is necessary to fix the orbital period, for which we adopt
the maximum \emph{a posteriori} period from the global fit. Additionally, we fix
the two quadratic limb darkening coefficients to values interpolated from
the \citet{claret:2011} tabulation of \emph{Kepler} bandpass coefficients as
generated from a PHOENIX stellar atmosphere model. The interpolation is made
at the point $T_{\mathrm{eff}} = 5308$\,K and $\log g = 4.632$, giving
$u_1 = 0.542$ and $u_2 = 0.139$. Fixing the limb darkening coefficients in these
fits imposes the condition that the host star is the same between the two
transits, but does not impose that the transiting body is the same.

We regress the standard \citet{mandel:2002} model to each transit using \multi\ 
and four free parameters, $\tau$, $p$, $\rho_{\star}$ and $b$, where we
use the same priors as before, except $\rho_{\star}$ changes to an uninformative
Jeffrey's prior. We also still enforce the blend factor, $B$, but now simply
fix it at the maximum likelihood value rather than imposing a Gaussian prior.
As with the limb darkening, this both simplifies the regression and enforces the
condition that the star is the same between the two transits.

In Table~\ref{tab:indivparams}, we report the maximum \emph{a posteriori} values 
for these parameters for each epoch, although we replace $b$ and $\rho_{\star}$ 
with the more intuitive transit durations $T_{14}$ and $T_{23}$. From 
Table~\ref{tab:indivparams}, it is evident that the two transits are consistent 
with being due to the same underlying transiting body, supporting our hypothesis 
made earlier in \S\ref{sub:period}. Also note that there is no evidence for 
precession effects between these two events.

\begin{table}
\caption{
Comparison of the four basic parameters describing a transit when epochs 1 and
2 are fitted independently.
} 
\centering 
\begin{tabular}{c c c} 
\hline\hline
Parameter & Epoch 1 & Epoch 2 \\ [0.5ex] 
\hline
$\tau$ [BKJD$_{\mathrm{UTC}}$-2,455,000] \dotfill & $325.7674_{-0.0012}^{+0.0012}$ & $1029.9658_{-0.0012}^{+0.0012}$ \\
$(R_P/R_{\star})$ \dotfill & $0.04990_{-0.00044}^{+0.00103}$ & $0.05057_{-0.00043}^{+0.00092}$ \\
$T_{14}$\,[hours] \dotfill & $15.748_{-0.094}^{0.128}$ & $15.774_{-0.089}^{+0.119}$ \\
$T_{23}$\,[hours] \dotfill & $14.122_{-0.171}^{+0.097}$ & $14.141_{-0.154}^{+0.090}$ \\ [1ex]
\hline 
\end{tabular}
\label{tab:indivparams} 
\end{table}

\subsection{Exomoon Fits}

The long-period nature of \kepx\ makes an exomoon search provocative, despite
the paucity of transit observations. In general, searching for an exomoon around 
a planet with just a few transits cannot yield a comprehensive search, since 
during these two events a moon of any size could happen not to transit the star. 
For this reason, upper limits on a moon's size are generally very large. With 
just two transits, deviations from a linear ephemeris cannot be detected and 
thus the transit timing variations (TTV) effect is lost, which is the easiest 
way to infer a moon's mass \citep{kipping:2009a,kipping:2009b}. Nevertheless, 
weaker constraints on an exomoon's mass can be inferred by transit duration 
variations, TDVs \citep{kipping:2009a,kipping:2009b}, and ingress/egress 
asymmetry \citep{luna:2011}. For this reason, a photodynamic algorithm
is the most reasonable way to model potential exomoons.

We regressed the analytic photodynamic \luna\ transit model \citep{luna:2011} of 
a planet with a single moon using the \multi\ algorithm. We adopt the same 
proceedures and priors as used in previous papers from the ``Hunt for Exomoons 
with Kepler'' (HEK) project (e.g. see \citealt{hek:2012,hek:2013,
kepler22:2013}).

\begin{figure*}
\begin{center}
\includegraphics[width=18.0 cm]{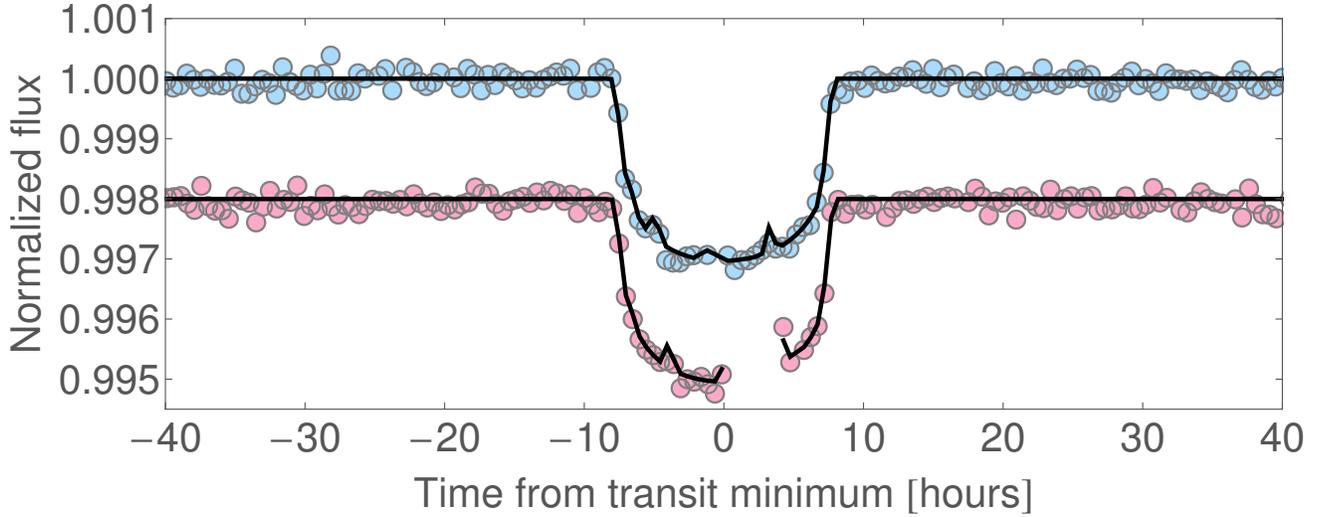}
\caption{Maximum a posteriori planet-with-moon light curve model generated
using \luna\ \citep{luna:2011} for the \kepx\ photometry. Blue points show the 
first transit epoch and red the second (vertically offset by 0.002). We argue 
that the depicted moon model is likely fitting out starspot crossings.} 
\label{fig:moonfits}
\end{center}
\end{figure*}

By comparing the Bayesian evidences, we find that the planet-with-moon model is 
favored over the simple planet-only model at 8.3\,$\sigma$. However, this
significance should not be used in isolation to claim moon detections, as
has been cautioned in previous HEK papers. Inspecting the posterior 
distributions for the 15 parameters in our model (14 usual planet-with-moon 
parameters used by the HEK project plus one extra for the blend factor $B$) 
reveals a preference for a short-period, close-in moon with $a_{SP}/R_P = 
3.52_{-0.57}^{+0.73}$ ($a_{SP}=$semi-major axis of the moon around the planet) 
and $P_S = 0.339_{-0.007}^{+0.0129}$\,days\footnote{These 
two terms can be converted into a mean density for the planet 
\citep{weigh:2011}, which is enforced to be physically plausible in our moon 
fits} (orbital period of the moon). This is apparent by plotting the maximum 
\emph{a posteriori} light curve model shown in Figure~\ref{fig:moonfits}, where 
the model produces mutual events (i.e. the moon and planet eclipse during a 
transit) at several locations, which is the dominant type of photometric effect 
of close-in moons \citep{luna:2011}.

The light curve features due to mutual events have a close morphological
resemblance to starspot crossings (e.g. see \citealt{pont:2007,rabus:2009}).
The features shown in Figure~\ref{fig:moonfits} have an amplitude 
$\lesssim300$\,ppm, which appears considerably larger than the rotational
modulation ampltidue shown later in \S\ref{sub:rotation}. However, the
rotational modulations may be caused by a mixture of dark and bright spots 
leading to an attenuated disk-integrated signal. We therefore consider the
hypothesis of starspot crossings to be a viable explanation for the 
features fitted out by the planet-with-moon model. Without unambiguous evidence
for what would be the first observation of a transiting exomoon, the model
should be considered to be not the favored hypothesis for this data. We compute
a 95\% confidence upper limit on the satellite-to-planet mass ratio of 
$(M_S/M_P)<0.26$, using our marginalized posterior distributions from the
photodynamic fit.

\section{DISCUSSION}
\label{sec:discussion}


\subsection{Insolation}
\label{sub:insolation}

As the longest period known transiting planet, \kepx\ should be much cooler than 
the typical transiting exoplanet. The orbit-averaged insolation received by 
\kepx, $S_{\mathrm{eff}}$, is

\begin{align}
\frac{S_{\mathrm{eff}}}{S_{\oplus}} &= \frac{(L_{\star}/L_{\odot})}{(a_P/\mathrm{AU})^2 \sqrt{1-e^2}},
\label{eqn:Seff}
\end{align}

where $S_{\oplus}$ denotes the orbit-averaged insolation received by the Earth.
Using our joint posterior distributions for the planetary and stellar 
parameters, $S_{\mathrm{eff}} = 0.276_{-0.028}^{+0.055}$\,$S_{\oplus}$. The 
equilibrium temperature of \kepx\ is then $T_P=202.0_{-5.3}^{+9.5}$\,K for a 
Bond albedo of zero, or $T_P=184.8_{-4.8}^{+8.6}$\,K using a more realistic 
assumption of a Uranian-like albedo of 0.30. 

With a semi-major axis of $1.22_{-0.11}^{+0.09}$\,AU, \kepx\ orbits closer to 
its parent star than the orbit of Mars (1.52~AU) around the Sun. Despite this 
smaller orbit, the lower luminosity of \kep\ ($L_{\star} = 
0.40\pm0.06$\,$L_{\odot}$) causes \kepx\ to receive just $\sim$64\%
of the insolation received by Mars (0.43\,$S_{\oplus}$). Comparing the incident 
insolation to the habitable-zone 
boundaries of \citet{kopparapu:2013}, \kepx\ lies firmly outside the maximum 
greenhouse outer edge.  Statistically, 79\% of the joint posterior samples lie 
beyond the maximum greenhouse outer boundary.

Due to the lower luminosity of \kep, we seek an alternative insolation-weighted
metric for comparing the orbit of \kepx\ with the planetary orbits in the Solar 
System. We define the ``effective semi-major axis'' as the semi-major axis of a 
circular orbit around the Sun where the exoplanet would receive the same 
insolation as in its current orbit around its host host. Mathematically, we have

\begin{align}
\frac{a_{\mathrm{eff}}}{\mathrm{AU}} &= \sqrt{ \frac{1}{ (S_{\mathrm{eff}}/S_{\oplus}) } },
\label{eqn:aeff}
\end{align}

where $S_{\mathrm{eff}}$ was defined earlier in Equation~\ref{eqn:Seff}. With 
this relation, the effective semi-major axis of \kepx\ is 
$1.90_{-0.17}^{+0.10}$\,AU. Thus, \kepx\ has an insolation roughly midway 
between the insolations of Mars and the asteroid Vesta ($a$ = 2.36~AU). 

\subsection{A Transiting Planet Near the Snow-Line}
\label{sub:snowline}

To place the insolation results in the context of planet formation theories, it 
is useful to compare $a_{\mathrm{eff}}$ with the location of the snow-line. The 
snow-line is an annulus in a protoplanetary disk where water ice condenses out 
of the gas \citep{sasselov:2000}. As the central star approaches the main 
sequence, time evolution of the disk temperature changes the location of the 
snow line \citep{kennedy:2006,garaud:2007,kennedy:2008}. In most planet 
formation theories, rocky terrestrial planets form inside the snow line; icy 
planets grow outside the snow line \citep{ida:2005}. 

In the Solar System, evidence from the asteroid belt suggests a ``canonical'' 
snow-line distance of around 2.7\,AU \citep{abe:2000,morbidelli:2000,
rivkin:2002}. However, the location in a general protoplanetary disk depends 
upon the luminosity of the central star and the grain opacities, mass accretion 
rates, and surface densities in the disk \citep{lecar:2006}. Time variations in 
these quantities change the position of the snow-line. Theoretical calculations 
of static protoplanetary disks suggest snow-line distances of 1.0--1.8~AU
\citep{sasselov:2000,lecar:2006}. Time-dependent calculations yield distances of
3~AU at 0.3~Myr to 1~AU at 10~Myr \citep{kennedy:2008}.  Our effective 
semi-major axis of $1.90_{-0.17}^{+0.10}$\,AU places \kepx\ beyond the snow-line 
for most of the evolution of the protosolar nebula.

\begin{figure}
\begin{center}
\includegraphics[width=8.4 cm]{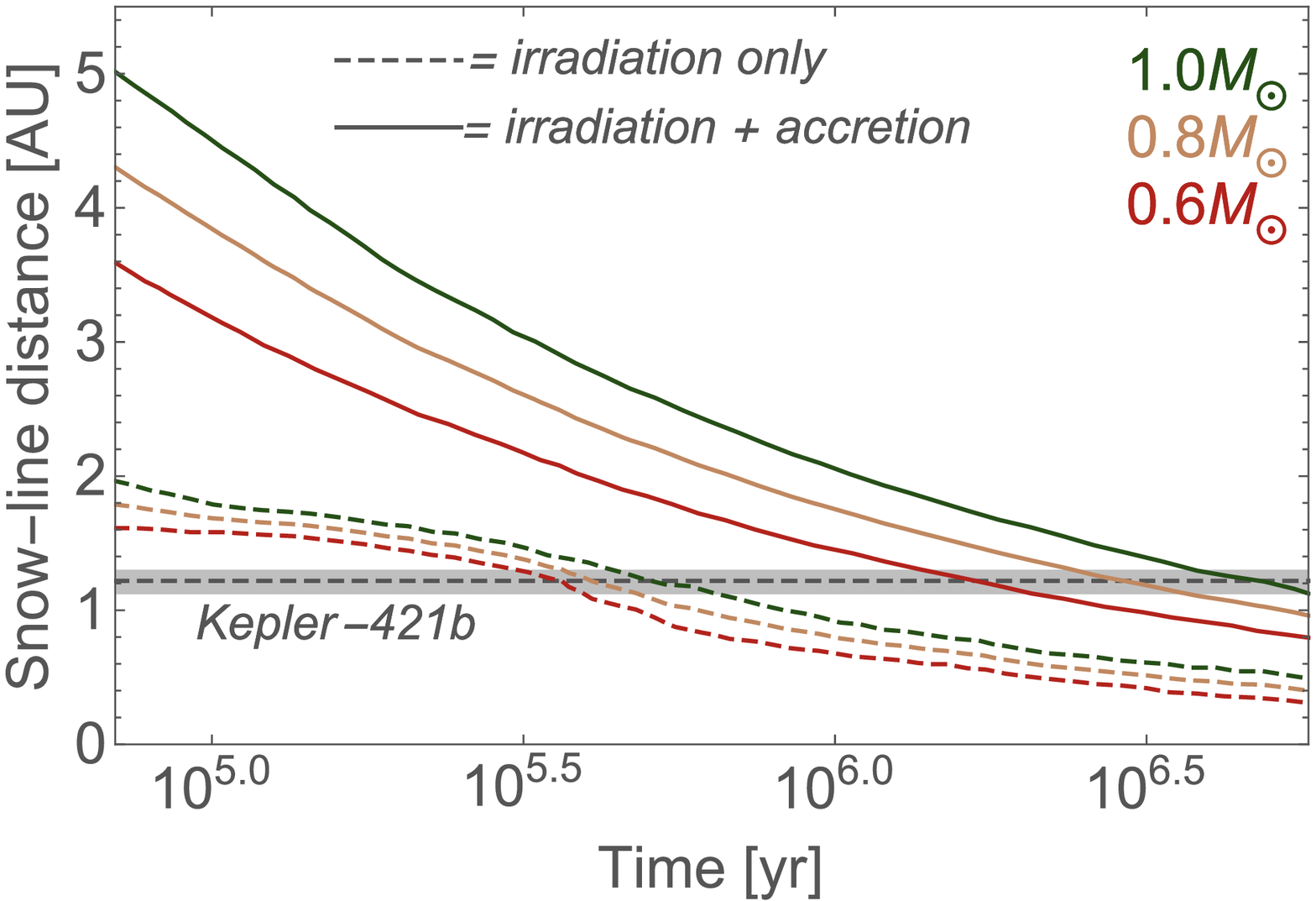}
\caption{
Location of the snow-line over time for a $0.6$\,$M_{\odot}$, $0.8$\,$M_{\odot}$
and $1.0$\,$M_{\odot}$ star, with irradiation only (using \citealt{palla:1999} 
PMS tracks, dashed line) and irradiation + accretion (solid line). Calculated
contours come from \citet{kennedy:2006}, except for the $0.8$\,$M_{\odot}$
contour which is an interpolation between the other two. The 68.3\% confidence
interval for \kepx's present location (gray band) equates to the snow-line
distance at $\sim3$\,Myr. 
}
\label{fig:snowline}
\end{center}
\end{figure}

To compare the actual semi-major axis of \kepx\ to protoplanetary disk models, 
we consider evolving disks around a 0.8\,$M_{\odot}$ star 
\citep[e.g.,][]{kennedy:2008}. The \citet{kennedy:2008} calculations derive the 
position of the snow line as a function of time in response to changes in the 
disk accretion rate and the stellar luminosity. We digitized the temporal 
evolution curves shown in their Figure~1 and then linearly interpolated between 
the 0.6\,$M_{\odot}$ and 1.0\,$M_{\odot}$ curves to estimate the location of 
\kep's snow-line over time, which is shown in Figure~\ref{fig:snowline}.

Using our posterior samples for the semi-major axis of \kepx\ (i.e. $a$ and not
$a_{\mathrm{eff}}$), the orbit of \kepx\ lies inside the snow-line for stellar 
ages exceeding $\tau\simeq2.9_{-0.5}^{+0.9}$\,Myr. This age is comparable to the
median lifetimes of protoplanetary disks around solar-type stars
\citep[e.g.,][]{strom:1993,haisch:2001}. With disk lifetimes scaling as 
$M_{\star}^{-1/2}$ \citep{yasui:2012}, it is quite feasible that \kep's 
protoplanetary disk remained at this time.

Snow-line transiting planets like \kepx\ may be common but their
discovery is impinged by the low transit probability ($\sim0.3$\%) and
number of events. \citet{wright:2009} highlighted an enhancement in the 
occurrence of exoplanets $\sim1$\,AU, which \citet{mordasini:2012} interpret as 
a signature of the snow-line. \citet{rice:2013} find that by normalizing 
exoplanet semi-major axes by the snow-line distance, there is evidence for a 
pile-up of planets around the snow-line threshold, suggesting \kepx-like worlds 
may be common.

\subsection{Formation Scenarios}
\label{sub:formation}

Although calculating detailed formation scenarios for \kepx\ is outside the
scope of this work, simple arguments suggest \kepx\ is an icy planet which
formed at or beyond the snow line. With a radius of roughly 4\,$R_{\oplus}$ and 
a mass density of at least 5\,g\,cm$^{-3}$, a rocky \kepx\ has a mass of at 
least 60\,$M_{\oplus}$. Growing such a massive planet requires a massive 
protostellar disk with most of the solid material at 1--2~AU \citep{mann:2010,
hansen:2012}. Among protoplanetary disks in nearby star-forming regions, such 
massive disks are rare \citep{andrews:2013}. Thus, a rocky \kepx\ seems 
unlikely.

Outside the snow-line, icy planets with radii of roughly 4\,$R_{\oplus}$ can 
form in more common, much less massive disks. In the standard core accretion 
theory, icy planets grow from the agglomeration of smaller planetesimals 
\citep[e.g.,][]{youdin:2013}. Depending on the initial sizes of the 
planetesimals growth times range from a few Myr to several Gyr \citep{mann:2010,
bromley:2011,kobayashi:2011,rogers:2011,lambrechts:2012,kenyon:2014}. As icy 
planets grow, they migrate through the gas or leftover planetesimals 
\citep{ida:2005,mann:2010,kenyon:2014}. Migrating planets sometimes reach small 
semi-major axes, $a\lesssim0.5$\,AU \citep{mann:2010,rogers:2011}. Often, the 
low surface densities of leftover planetesimals preclude migration inside 
1--2\,AU \citep{kenyon:2014}.

For \kepx, {\it in situ} formation is a reasonable alternative to formation and
migration from larger semi-major axes. Scaling results from published 
calculations, the time scale to produce a 10--20\,$M_{\oplus}$ planet is 
comparable to or larger than the median lifetime of the protoplanetary disk 
\citep[e.g.,][]{mann:2010,rogers:2011,lambrechts:2012,kenyon:2014}. Thus, 
formation from icy planetesimals is very likely. If significant migration 
through the gas \citep[e.g.,][]{lega:2014} and leftover planetesimals 
\citep[e.g.,][]{kenyon:2014} can be avoided, \kepx\ remains close to the 
``feeding zone'' in which it formed.

Developing a more rigorous theory requires more information 
\citep[see the discussion in][]{kenyon:2014}. Good estimates for the mean 
density can distinguish between rocky and icy formation scenarios. Detecting 
absorption from atmospheric constituents might enable choices between various 
migration scenarios.


\subsection{Future Follow-Up}

Within the \emph{Kepler} time series, only two transits of \kepx\ are observed.
The expected third transit would have occurred in March 2014, after the primary 
\emph{Kepler Mission} ended. Unfortunately, we did not have time to schedule 
observations capable of detecting the event. The fourth transit is due at 
$2,457,438.3627_{-0.0041}^{+0.0042}$\,$\mathrm{BJD}_{\mathrm{UTC}}$, which is in
February 2016. Note that these estimates assume a strictly linear ephemeris,
for which we have no direct evidence given that only two transits have been
observed thus far.

Assuming a Uranian mean density for \kepx, we estimate that the radial velocity 
semi-amplitude induced on \kep\ would be $K = 1.40_{-0.14}^{+0.20}$\,m/s with
a periodicity of 704\,d. This clearly presents a significant challenge to
current observational facilities, but \kepx\ is a unique object being the first
Neptune-like planet discovered at long-period by transits or radial velocity.
Determining the mass of the first transiting cold-Neptune would provide a
crucial point in intepretting the empirical mass-radius relationship of
exoplanets as a function of insolation.

\acknowledgements
\section*{Acknowledgements}

This work made use of the Michael Dodds Computing Facility and the Pleiades
supercomputer at NASA Ames.
This work was performed [in part] under contract with the California Institute 
of Technology (Caltech)/Jet Propulsion Laboratory (JPL) funded by NASA through 
the Sagan Fellowship Program executed by the NASA Exoplanet Science Institute.
GT acknowledges partial support for this work from NASA grant NNX14AB83G 
(\kepler\ Participating Scientist Program).
We offer our thanks and praise to the extraordinary scientists, engineers
and individuals who have made the \emph{Kepler Mission} possible.
We also thank C.~Burke and J.~Twicken for assistance in obtaining the centroid
motion results.
%


%

\end{document}